\documentclass{llncs}

\title{Unique Pattern Matching in Strings}
\author{Stijn Vansummeren\thanks{Research Assistant of the Fund for
     Scientific Research - Flanders (Belgium) }}
 \institute{University of Limburg - Belgium\\
   \email{stijn.vansummeren@luc.ac.be}}

\newcommand{\emptyseq}{\ensuremath{\varepsilon}}
\newcommand{\emptyw}{\ensuremath{\lambda}}
\newcommand{\con}{\ensuremath{\cdot}}
\newcommand{\cor}{\ensuremath{+}}

\newcommand{\labs}{{\Sigma}} 
\newcommand{\labsc}{{\Sigma_{\cm}}} 
\newcommand{\nat}{\mathbb{N}}

\newcommand{\cm}{\ensuremath{\Box}}
\newcommand{\ndef}{\ensuremath{\perp}}

\newcommand{\w}{\ensuremath{w}}
\newcommand{\pat}{\texttt{P}}

\newcommand{\dom}{\ensuremath{\mathrm{dom}}} 
\newcommand{\nodedom}{\ensuremath{\{1,2\}^*}} 
\newcommand{\bn}[1]{\ensuremath{\mathrm{bn}(#1)}} 

\newcommand{\typeof}[3]{\ensuremath{\mathcal{T}(#1,#2,#3)}}

\newcommand{\match}[3]{\ensuremath{{#1} \in {#2} \rightsquigarrow {#3}}}
\newcommand{\br}[3]{\ensuremath{M(#1,#2,#3)}}
\newcommand{\ipi}{\ensuremath{\pi^{-1}}} 

\newcommand{\width}[1]{\ensuremath{|#1|}}
\newcommand{\pos}[2]{\ensuremath{\mathrm{pos}(#1,#2)}}
\newcommand{\bound}[2]{\ensuremath{#1|_{#2}}}

\usepackage{amsmath}
\usepackage{amssymb}
\usepackage{mathpartir}
\usepackage{pstricks,pst-node,pst-text, pst-tree}
\usepackage{algorithm, algorithmic}

\begin{document}
\maketitle
\begin{abstract}
Regular expression patterns are a key feature of document processing
languages like Perl and XDuce.  It is in this context that the first
and longest match policies have been proposed to disambiguate the
pattern matching process. We formally define a matching semantics with
these policies and show that the generally accepted method of
simulating longest match by first match and recursion is incorrect. We
continue by solving the associated type inference problem, which
consists in calculating for every subexpression the set of words the
subexpression can still match when these policies are in effect, and
show how this algorithm can be used to efficiently implement the
matching process.
\end{abstract}

\newpage

\section{Introduction}
\label{sec:introduction}

Using regular (tree) expression patterns to extract relevant data from
a string (or tree) is a highly desirable feature for programming
languages supporting document transformation or data retrieval.
Indeed, it is a core feature of Perl \cite{perl} and has recently been
proposed in the context of the XML programming language XDuce
\cite{xduce:matching,xduce:thesis}. The matching process consists of
two parts: (1) ensuring that the input belongs to the language of the
expression; and, (2) associating with every subexpression the matching
part of the input.  In general, patterns can be ambiguous, meaning
that there are various ways of matching the input, resulting in
multiple associations. When regular expression patterns are used as
database queries, it is indeed common and desirable for a pattern to
have many matches in the data, and to be able to retrieve all of them.
However, in general-purpose programming using pattern matching as in
ML, Prolog, or XDuce, we normally want unique matching and a
deterministic semantics.

One approach to the latter problem would be to simply disallow
ambiguity by requiring the regular expressions to be
\emph{unambiguous} \cite{book}.  Another, more programmer-friendly
approach is to allow arbitrary regular expression patterns, but to
employ a special \emph{unique matching} semantics.  In this paper, we
investigate this last approach on strings.  We present a formal
definition of unique matching, and give a sound and complete algorithm
for solving the associated \emph{regular type inference problem}:
given a regular expression $\pat$ and a regular ``context language''
$C$, compute for each subexpression $\pat'$ of $\pat$ the regular
language consisting of all subwords $\w'$ of an input string $\w \in
C$, such that $\w'$ is matched by $\pat'$ when matching $\w$ uniquely
to $\pat$.

Regular type inference is useful for type-checking transformations:
given an input language, does the transformed document always adhere
to a desired output language
\cite{suciu,lambdare,xduce:language,xduce:regtypes}? An important
feature of our approach is that it directly yields an unambiguous NFA
that not only contains the types of all the subexpressions of the
given pattern, but also serves to perform the actual matching on any
given string in linear time.

Regular expression pattern matching and its type inference problem was
first studied in the context of XDuce, an XML processing language
\cite{xduce:matching}. While enormously influential, it suffers from a
few disadvantages.  First, the XDuce type inference algorithm is
incomplete.  Also, the formalism used to represent regular tree
languages (in terms of linear context-free grammars with encoding into
binary trees) is hardwired into the algorithm, making it very
syntactic in nature and hard to understand.  Our aim is to abstract
away from a particular syntax of regular languages. We therefore
present a sound and complete algorithm using only operations on
languages.  As such, our algorithm is independent of a surrounding
(regular expression) type system.

Another problem with the XDuce approach is the introduction of a
misconception regarding unique matching of regular expression
patterns. Namely, the longest match policy used to disambiguate the
Kleene closure is simulated by recursion and the first match policy,
which is used to disambiguate disjunctions. We will show that this
simulation is incorrect.

Two recent followups on XDuce, which happened concurrently and
independently with our own work, are $\mathbb{C}$Duce and
$\lambda^{\mathrm{re}}$ \cite{lambdare,cduce,subtyp}. While both
approaches claim complete type inference, they follow XDuce in
simulating longest match by first match and recursion. We will show
that this causes the inferred types to be incorrect with regard to the
longest match policy. Another advantage of our approach is the
elementary nature of our type inference method, which works purely on
the language level and which yields to a reasonably simple correctness
proof.

The rest of this paper is organized as follows. In Sect.
\ref{sec:patterns} we formally define the matching relation based on
two disambiguating rules: first match and longest match. It is shown
that the above-mentioned simulation of longest match by first match
and recursion is incorrect. In Sect. \ref{sec:typeinfer} we introduce
the type inference problem and give a declarative way of solving it. A
concrete implementation strategy is given in Sect. \ref{sec:unifying},
where we also show that this strategy leads to an efficient
implementation of the matching process. The last section touches on
some future work.

\section{Unique Pattern Matching}
\label{sec:patterns}

Matching against regular expression patterns is done in two parts: (1)
making sure that the input word belongs to the language of the
expression; and, (2)~associating with every subexpression the matching
subword.  These associations can then be used to extract relevant data
from the matched string. However, regular expressions can be ambiguous
\cite{book}. For instance, when we want to show that $aa$ is matched
by $a^* \con (a \cor \emptyseq)$, should we associate $aa$ to $a^*$
and $\emptyw$ to $(a \cor \emptyseq)$ or should we associate $a$ to
$a^*$ and $a$ to $(a \cor \emptyseq)$?\footnote{To avoid confusion we
  denote the empty word with $\emptyw$ and the regular expression
  recognizing $\emptyw$ with $\emptyseq$.} And what if we consider $a$
being matched by $a \cor a$, should we associate $a$ with the first
$a$ or the second?  Furthermore, how should we deal with the matching
of $aab$ by $(a \cor b \cor a\con b)^*$? In order to get a unique
matching strategy, we define the $*$-operator to be ``greedy'',
meaning it should match the longest possible subword still allowing
the rest of the pattern to match. This is referred to as the
\emph{longest match} semantics in Perl, XDuce, $\mathbb{C}$Duce and
$\lambda^{\mathrm{re}}$. Furthermore, for patterns like $\pat_1 \cor
\pat_2$, we opt for the \emph{first match} policy where we only
associate the subword with $\pat_2$ if it cannot be matched by
$\pat_1$.  Finally, we treat $\pat^*$ as being atomic, in the sense
that we do not give associations for subexpressions of $\pat$.  In
this section, we give formal definitions of patterns and the matching
process and show that the proposed policies guarantee a unique
matching strategy.

We assume to be given a fixed, finite alphabet $\labs$ which does not
contain the special symbols $\ndef$ and $\cm$. Elements of $\labs$
will be denoted with $\sigma$ and words over $\labs$ will be denoted
with $\w$ throughout the rest of this paper. A \emph{regular
  expression pattern} $\pat$ is a regular expression over $\labs$.
That is, $\pat$ is either of the form $\sigma$ with $\sigma \in
\labs$, $\pat_1 \cor \pat_2$, $\pat_1 \con \pat_2$ or $\pat_1^*$,
where $\pat_1$ and $\pat_2$ are already regular expression patterns.
We define all operators to be right-associative. The set of all
patterns is denoted by $\mathcal{P}$.  Because we will consider the
abstract syntax tree of a pattern, we abuse notation slightly and
identify $\pat$ with the partial function $\pat: \nodedom \rightarrow
\{*, \con, \cor, \emptyseq\} \cup \labs$ such that
\begin{itemize}
\item if $\pat = \emptyseq$ then $\dom(\pat) = \{\emptyw\}$ and
  $\pat(\emptyw) = \emptyseq$,
\item if $\pat = \sigma$ with $\sigma \in \labs$ then $\dom(\pat) =
  \{\emptyw\}$ and $\pat(\emptyw) = \sigma$,
\item if $\pat = \pat_1 \cor \pat_2$ then $\dom(\pat) = \{\emptyw\}
  \cup \{1n \mid n \in \dom(\pat_1)\} \cup \{2n \mid n \in \dom(\pat_2)\}$
  with $\pat(\emptyw) = \cor$, $\pat(1n) = \pat_1(n)$ and $\pat(2n) =
  \pat_2(n)$,
\item if $\pat = \pat_1 \con \pat_2$ we make a similar definition,
  only $\pat(\emptyw) = \con$, and
\item if $\pat = \pat_1^*$ then $\dom(\pat) = \{\emptyw\} \cup \{1n
  \mid n \in \dom(\pat_1)\}$, $\pat(\emptyw) = *$ and $\pat(1n) =
  \pat_1(n)$.
\end{itemize}
Intuitively the function view of a pattern describes the abstract
syntax tree of its regular expression, as shown in Fig.
\ref{fig:pattern}. Elements of \nodedom\ are called \emph{nodes} and
will be denoted by $n$, $m$ and their subscripted versions. We say
that node $n$ is an ancestor of $m$ if there is some $n' \not =
\emptyw$ for which $m = n n'$. If $m = n1$ ($m = n2$) then $m$ is the
left (right) child of $n$. A node $n \in \dom(\pat)$ is a
\emph{bindable node} of $\pat$ if it does not have an ancestor labeled
with $*$. The set of bindable nodes of $\pat$ is denoted with
$\bn{\pat}$. We define the size $\width{\pat}$ of $\pat$ as the
cardinality of its domain.

\newcommand{\ann}[2]{\nput[labelsep=.16]{#1}{\pssucc}{#2}}
\begin{figure}[tb]
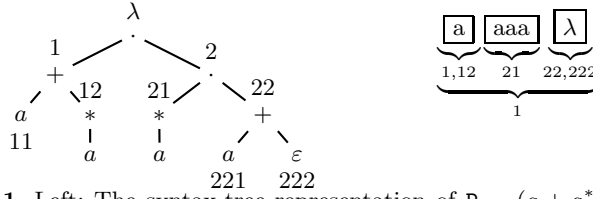

  \centering
  \begin{tabular}[h]{p{.4\linewidth}p{4ex}p{.4\linewidth}}
    \pstree[nodesep=3pt,levelsep=15pt]{%
      \TR[name=nemptyw]{$\con$}\ann{90}{$\emptyw$}}{%
      \pstree{\TR[name=n1]{$\cor$}\ann{90}{$1$}}{%
        \TR{$a$}\ann{-90}{$11$}
        \pstree{\TR[name=n12]{$*$}\ann{90}{$12$}}{
          \TR{$a$}
        }
      }
      \pstree{\TR{$\con$}\ann{90}{$2$}}{
        \pstree{\TR{$*$}\ann{90}{$21$}}{
          \TR{$a$}
        }
        \pstree{\TR{$\cor$}\ann{90}{$22$}}{
          \TR{$a$}\ann{-90}{$221$}
          \TR{$\emptyseq$}\ann{-90}{$222$}
        }
      }
    }
    &   &
        $ \underbrace{\underbrace{\fbox{a}}_{1,12}
          \underbrace{\fbox{aaa}}_{21}
          \underbrace{\fbox{\emptyw}}_{22,222}}_{1}$ 
   \end{tabular} 

  \caption{Left: The syntax-tree representation of $\pat = (a 
    \cor a^*) \con a^* \con (a \cor \emptyseq)$. The bindable nodes
    have their addresses annotated. Right: The associations resulting
    from the matching of $\pat$ against $aaaa$. Nodes that are not mentioned are associated with $\ndef$.}
  \label{fig:pattern}
\end{figure}
The matching process is formally described by the matching relation
$\match{\w}{\pat}{V}$, signifying that $\w$ is matched by $\pat$
yielding associations $V$. Here, $V$ is a function from $\bn{\pat}$ to
subwords of $\w$ or to the special symbol $\ndef$.  Intuitively, $V(n)
= \w'$ if the pattern rooted at node $n$ is responsible for matching
the subword $\w'$. It is $\ndef$ if the subpattern is not responsible
for recognizing any subword of $\w$.  To simplify the definition of
the matching relation we introduce some notation. If $V_1$ and $V_2$
are such functions, then we use $V_1 \cor \pat_2$ to denote the
function for which $(V_1 \cor \pat_2)(\emptyw) = V_1(\emptyw)$, $(V_1
\cor \pat_2)(1n) = V_1(n)$ for every $n \in \dom(V_1)$ and $(V_1\ 
\cor\ \pat_2)(2n) = \ndef$ for every $n \in \bn{\pat_2}$.  We define
$\pat_1 \cor V_2$ similarly. Moreover, $V_1 \con V_2$ is the function
such that $(V_1 \con V_2)(\emptyw) = V_1(\emptyw) \con V_2(\emptyw)$
if $V_1(\emptyw) \not = \ndef$ and $V_2(\emptyw) \not = \ndef$, and it
is $\ndef$ otherwise. Furthermore, $(V_1 \con V_2)(1n) = V_1(n)$ for
every $n \in \dom(V_1)$ and $(V_1 \con V_2)(2n) = V_2(n)$ for every $n
\in \dom(V_2)$.

The inference rules for $\match{\w}{\pat}{V}$ are given in Fig.
\ref{fig:matching}. We write $\w \in \pat$ if $\match{\w}{\pat}{V}$
holds for some $V$ and $\w \not \in \pat$ otherwise. The auxiliary
relation $\match{(\w_1, \w_2)}{\pat_1 \con \pat_2}{(V_1 , V_2)}$ is
used to indicate that when matching $\w_1 \w_2$ by $\pat_1 \con
\pat_2$, pattern $\pat_1$ is responsible for matching $\w_1$, yielding
associations $V_1$, while $\pat_2$ is responsible for matching $\w_2$,
yielding associations $V_2$. The first match policy is implemented in
rules Or2 and COr2 where we do not allow to examine the second branch
of a disjunction until the first one fails to match. The longest match
policy is expressed in rule CKleene. Figure \ref{fig:pattern} shows
the associations obtained by matching $(a \cor a^*) \con a^* \con (a
\cor \emptyseq)$ against $aaaa$.

\begin{figure}[tb]
  \centering
{\scriptsize
\begin{mathpar}
  \inferrule[Empty]{}{\match{\emptyw}{\emptyseq}{[\emptyw \rightarrow
      \emptyw ]}}
\and
\inferrule[Lab]{}
     {\match{\sigma}{\sigma}{[\emptyw \rightarrow \sigma]}}
\and
\inferrule[Kleene Empty]{}{\match{\emptyw}{\pat^*}{[\emptyw
    \rightarrow \emptyw]}}
\and
\inferrule[Kleene Closure]
     {\match{\w_1}{\pat}{V_1} \\
       \match{\w_2}{\pat^*}{V_2} \\
       \w_1 \not = \emptyw
     }
     {\match{\w_1 \w_2}{\pat^*}{[\emptyw \rightarrow w_1 w_2]}}
\and
\inferrule[Or1]
     {\match{\w}{\pat_1}{V}}
     {\match{\w}{\pat_1 \cor \pat_2}{V\cor \pat_2}}
\and
\inferrule[Or2]
     {\match{\w}{\pat_2}{V} \\ \w \not \in \pat_1}
     {\match{\w}{\pat_1 \cor \pat_2}{\pat_1 \cor V}}
\and

\inferrule[CElem]
     {\match{(\w_1,\w_2)}{\pat_1 \con \pat_2}{(V_1, V_2)}}
     {\match{\w_1 \w_2}{\pat_1\con \pat_2}{V_1 \con V_2}}
\and
\inferrule[CEmpty]
     {\match{\w}{\pat}{V_2} }
     {\match{(\emptyw, \w)}{\emptyseq \con \pat}{([\emptyw \rightarrow
     \emptyw],V_2)}}
\and
\inferrule[CLab]
     {\match{\sigma}{\sigma}{V_1} \\ \match{\w}{\pat}{V_2} }
     {\match{(\sigma, \w)}{\sigma \con \pat}{(V_1, V_2)}}
\and
\inferrule[COr1]
     {\match{(\w_1, \w_2)}{\pat_1 \con \pat_3}{(V_1, V_2)}}
     {\match{(\w_1, \w_2)}{(\pat_1 \cor \pat_2) \con \pat_3}{(V_1 
     \cor \pat_2,V_2)}}
\and
\inferrule[COr2]
     {\match{(\w_1, \w_2)}{\pat_2 \con \pat_3}{(V_1, V_2)} \\\\
       \w_1 \w_2 \not \in \pat_1 \con \pat_3}
     {\match{(\w_1, \w_2)}{(\pat_1 \cor \pat_2) \con \pat_3}{(\pat_1 
     \cor V_1,V_2)}}
\and
\inferrule[CCon]
     {\match{(\w_1, \w_2 w_3)}{\pat_1 \con (\pat_2 \con
     \pat_3)}{(V_1,W)} \\\\
      \match{(\w_2, \w_3)}{\pat_2 \con \pat_3}{(V_2, V_3)}}
     {\match{(\w_1 \w_2, \w_3)}{(\pat_1 \con \pat_2) \con
     \pat_3}{(V_1 \con V_2, V_3)}}
\and
\inferrule[CKleene]
     {\match{\w_1}{\pat_1^*}{V_1} \\ 
       \match{\w_2}{\pat_2}{V_2} \\\\ 
       \neg(\exists w_3 \not = \emptyw, \w_4: \w_2 = \w_3 \w_4 \wedge
     \w_1 \w_3 \in \pat_1^* \wedge \w_4 \in \pat_2)}
     {\match{(\w_1, \w_2)}{\pat_1^* \con \pat_2}{(V_1,V_2)}}
\end{mathpar}

}  
  \caption{The matching relation}
  \label{fig:matching}
\end{figure}

The \emph{language} $L(\pat)$ of a pattern $\pat$ is defined as the
language of its regular expression. We can then obtain the following
theorem: 

\begin{theorem} 
  \label{theorem:welldef}
  The matching relation of Fig. \ref{fig:matching} is well defined:
  \begin{enumerate}
  \item \label{it:correct} The matching relation is semantically correct:
    $\match{\w}{\pat}{V}$ iff $\w \in L(\pat)$, and,
  \item \label{it:unique} The matching relation is unique:
    if $\match{\w}{\pat}{V}$ and $\match{\w}{\pat}{W}$ then $V = W$.
  \end{enumerate}
\end{theorem}
\begin{proof}[Sketch]
  The ``if'' part of (\ref{it:correct}) can be proved by induction on
  the matching derivation. To prove the other way around, we first
  define the relation $\sqsupset\, \subseteq \mathcal{P} \times
  \mathcal{P}$ where $\sqsupset$ relates a pattern with its immediate
  sub-patterns if $\pat \not = (\pat_1 \con \pat_2) \con \pat_3$ and
  $\pat \not = (\pat_1 \cor \pat_2) \con \pat_3$. We define
  $\sqsupset$ to relate $(\pat_1 \con \pat_2) \con \pat_3$ with
  $\pat_2 \con \pat_3$ and with $\pat_1 \con (\pat_2 \con \pat_3)$ and
  to relate $(\pat_1 \cor\ \pat_2) \con \pat_3$ with $\pat_1 \con
  \pat_3$ and with $\pat_2 \con \pat_3$. The monotone embedding $\phi$
  into $\nat \times \nat$ where $\phi(\pat) = (\width{\pat}, 0)$ if
  $\pat \not = \pat_1 \con \pat_2$ and $\phi(\pat_1 \con \pat_2) =
  (\width{\pat_1 \con \pat_2}), \width{\pat_1})$ otherwise, shows that
  $\sqsupset$ is a well-founded ordering on $\mathcal{P}$.  The proof then
  goes by well-founded induction on $(\mathcal{P}, \sqsupset)$.
  Statement (\ref{it:unique}) is proved by induction on both matching
  derivations.\qed
\end{proof}

In related work \cite{xduce:matching,xduce:thesis,lambdare,cduce}, the
longest match policy is simulated by the first match policy and
recursion. We will now show that this simulation is incorrect.
Concretely, consider the inference rule CKleene' for Kleene Closure in
a concatenation from $\lambda^{\mathrm{re}}$ (XDuce and
$\mathbb{C}$Duce use equivalent rules).
\begin{mathpar}
  \scriptsize
  \inferrule[CKleene']
     {\match{(\w_1, \w_2)}{((\pat_1 \con \pat_1^*) \cor \emptyseq)
     \con \pat_2}{(V_1,V_2)}} 
     {\match{(\w_1, \w_2)}{\pat_1^* \con \pat_2}{(V_1,V_2)}}
\end{mathpar}
Here, we assume w.l.o.g. $\emptyw \not \in \pat_1$. The proposed
intuition behind this rule is that, when trying to derive
$\match{\w}{\pat_1^* \con \pat_2}{V}$, we will be forced by the first
match policy to consider $(\pat_1 \con \pat_1^*) \con \pat_2$ before
$\emptyseq \con \pat_2$ at every expansion of $\pat_1^* \con \pat_2$.
Since $\emptyw \not \in \pat_1$, this should require us to split $\w$
into $\w_1 \in \pat_1^*$ and $\w_2 \in \pat_2$ such that $\w_2$ is the
smallest suffix of $\w$ still matched by $\pat_2$. However, this is a
false intuition.  Indeed, because the first match strategy continues
to be used in $\pat_1$, it is possible that $\pat_2$ is allowed to
start matching before a longer matching alternative in $\pat_1$ is
considered.  For example, consider the matching of $ab$ against $\pat
= (a \cor a \con b)^* \con (b \cor \emptyseq)$. By the longest match
policy, we would expect $(a \cor a \con b)^*$ to be associated with
$ab$.  Indeed, this is the unique association derived by our rules of
Fig.~\ref{fig:matching}.  Rule CKleene', however, will incorrectly
derive the association of $a$ to $(a+a\cdot b)^*$ and $b$ to
$(b+\varepsilon)$, as we show next.

\begin{mathpar}
  \scriptsize
  \inferrule*[Right=CElem]
  { 
    \inferrule*[Right=CKleene']{
      \inferrule*[Right=COr1]
      {
        \inferrule*[Right=CCon]
        {
          {
            \match{(a,b)}{(a \cor a \con b) \con ((a \cor a \con
              b)^*) \con (b \cor \emptyseq))}{(V'_1,W)}
          }\\
          {
            \match{(\emptyw,b)}{((a \cor a \con b)^*) \con (b \cor
              \emptyseq)}{(V'_2,V_2)}
          }
        }
        {
          \match{(a,b)}{((a \cor a \con b) \con (a \cor a \con b)^*)
            \con (b \cor \emptyseq)}{(V_1 = V'_1 \con V'_2, V_2)}
        }
      }
      {
        \match{(a,b)}{(((a \cor a \con b) \con (a \cor a \con b)^*) \cor
          \emptyseq) \con (b \cor \emptyseq)}{(V_1,V_2)}
      }
    }
    {
      \match{(a, b)}{(a \cor a \con b)^* \con (b \cor \emptyseq)}{(V_1, V_2)}
    }
  }
  {\match{ab}{\pat}{V_1 \con V_2} }
\end{mathpar}
The derivation for the first subgoal $\match{(a,b)}{(a \cor a \con b)
  \con ((a \cor a \con b)^*) \con (b \cor \emptyseq))}{(V'_1,W)}$
must look like
\begin{mathpar}
  \scriptsize
  \inferrule*[Right=COr1]
  {
    \inferrule*[Right=CLab]
    {
      \inferrule*[Right=Lab]
      { }
      {\match{a}{a}{V''_1 = [\emptyw \rightarrow a]}}
      \\
      \inferrule*[Right=CElem]
      { \dots }
      {\match{b}{( a \cor a \con b)^* \con (b \cor \emptyseq)}{W}
      }
    }
    {
      \match{(a,b)}{a \con (( a \cor a \con b)^* \con (b \cor
        \emptyseq))}{(V''_1, W)} 
    }
  }
  {
    \match{(a,b)}{(a \cor a \con b) \con (( a \cor a \con b)^* \con (b \cor
      \emptyseq))}{(V'_1 = V''_1 \cor (a \con b), W)} 
  }
\end{mathpar}
Here, the subderivation indicated by the dots above the use of CElem
is isomorphic to the derivation for the second subgoal
$\match{(\emptyw,b)}{( a \cor ab)^* \con (b \cor \emptyseq)}{(V'_2,
  V_2)}$:

\begin{mathpar}
  \scriptsize
  \inferrule*[Right=CKleene']
  {
    \inferrule*[Right=COr2]
    {
      \inferrule*[Right=CEmpty]
      {
        \inferrule*[Right=Or1]
        {
          \inferrule*[Right=Lab]
          { }
          {\match{b}{b}{V_2 = [\emptyw \rightarrow b]}}
        }
        {
          \match{b}{(b \cor \emptyseq)}{V_2}
        }
      }
      {
        \match{(\emptyw, b)}{\emptyseq \con (b \cor
          \emptyseq)}{(V'_2 = [\emptyw \rightarrow \emptyw], V_2)}
      }\\
      b \not \in ((a \cor a\con b) \con (a \cor a \con b)^*) \con (b
      \cor \emptyseq)
    }
    {
      \match{(\emptyw, b)}{(((a \cor a \con b) \con (a \cor a \con b)^*)
        \cor \emptyseq) \con (b \cor \emptyseq)}{(V'_2,V_2)}      
    }
  }
  {
    \match{(\emptyw,b)}{( a \cor a \con b)^* \con (b \cor \emptyseq)}{(V'_2, V_2)}
  }
\end{mathpar}
Now $(V_1 \con V_2)(1) = a$ and $(V_1 \con V_2)(2) = b$, as we wanted to
show.

\section{Type Inference}
\label{sec:typeinfer}
The matching process described in the previous section is used in many
practical languages (including Perl) where the associations
are used to construct the output.  Recently, there has been growing
interest to add \emph{type safety} to such languages: given an input
language, the transformation should always produce outputs adhering to
a certain output language \cite{suciu,lambdare,xduce:language}. In one
approach to achieving this, one has to infer for every subexpression
in the pattern the set of words it is capable of matching. In this
section we present an algorithm for this type inference problem. We
also note that the existing type inference algorithms are incorrect
with regard to the longest match policy; this will follow from the
incorrect simulation of longest match by recursion and first match, as
shown in the previous section.

Let $C$ be a set of words called the \emph{context}.  The \emph{type}
of a bindable node $n$ in $\pat$ relative to $C$, denoted as
$\typeof{n}{\pat}{C}$ is the set of words $\w$ for which there exists
some $\w' \in C$ such that $\match{\w'}{\pat}{V}$ and $V(n) = \w$. The
main result of this paper can be stated as follows:

\begin{theorem}
  \label{theorem:inferalgo}
  If $C$ is a regular language then $\typeof{n}{\pat}{C}$ is also
  regular, and can be effectively computed.
\end{theorem}

The algorithm is obtained by structural induction on the pattern,
applying the equalities we introduce in the following lemmas and
propositions.\footnote{Proofs of the claims in this section are given in
  Appendix~\ref{app:proofinfer}.} For instance, the next lemma shows
how to calculate $\typeof{\emptyw}{\pat}{C}$:
\begin{lemma}
  \label{lemma:toptype}
  $\typeof{\emptyw}{\pat}{C} =  L(\pat) \cap C$ for any pattern $\pat$.
\end{lemma}
\begin{proof}
  By a simple induction on the matching derivation we can prove that
  if $\match{\w}{\pat}{V}$ then $V(\emptyw) = \w$ and if
  $\match{(\w_1, \w_2)}{\pat_1 \con \pat_2}{(V_1, V_2)}$ then
  $V_1(\emptyw) = \w_1$ and $V_2(\emptyw) = \w_2$. Combining this
  observation with Theorem \ref{theorem:welldef}
  gives the desired result almost immediately . \qed
\end{proof}
Note that this result completely solves the type inference problem
when $\pat$ equals $\emptyseq$, $\sigma$ or $\pat_1^*$, since
$\bn{\pat} = \{\emptyw\}$ in these cases. When $\pat$ is of a
different form, we will calculate $\typeof{n}{\pat}{C}$ from
$\typeof{n'}{\pat'}{C'}$ for some simpler pattern $\pat'$ and possibly
different context $C'$. In the case where $\pat = (\pat_1 \con \pat_2)
\con \pat_3$ we will need the set $\br{n}{\pat}{C} = \{ \w_1 \cm \w_2
\mid \exists \w' \in C, \match{\w'}{\pat}{V}, V(n1) = \w_1, V(n2) =
\w_2\}$ to be defined on all bindable nodes of $\pat$ which are
labeled with a concatenation.  We will also show how to calculate this
set.  Intuitively, the symbol $\cm$ specifies how $\w_1\w_2$ is
``broken up'' into subwords when it is matched by the concatenation at
node $n$.

The case where $\pat = \pat_1 \cor \pat_2$ is handled by the following
proposition:

\begin{proposition} 
  \label{prop:or}
  For $\pat = \pat_1 \cor \pat_2$, the following equalities hold:
  \begin{enumerate}
  \item $\typeof{\emptyw}{\pat}{C} = \typeof{\emptyw}{\pat_1}{C} \cup
  \typeof{\emptyw}{\pat_2}{C - L(\pat_1)}$
  \item $\typeof{1n}{\pat}{C} = \typeof{n}{\pat_1}{C}$
  \item $\typeof{2n}{\pat}{C} = \typeof{n}{\pat_2}{C - L(\pat_1)}$
  \item $\br{1n}{\pat}{C} = \br{n}{\pat_1}{C}$
  \item $\br{2n}{\pat}{C} = \br{n}{\pat_2}{C - L(\pat_1)}$
  \end{enumerate}
\end{proposition}
The next two propositions handle $\pat = \emptyseq \con \pat_2$
and $\pat = \sigma \con \pat_2$. Here, the left quotient of
language $L$ by language $K$, defined as $\{s \mid \exists p \in K: ps
\in L\}$, is denoted as $K \backslash L$ . The \emph{right quotient} of
$L$ by $K$, defined as $\{p \mid \exists s \in K: ps \in L\}$,
is denoted as $L/K$. It is well-known that regular languages
are closed under both quotients \cite{ullman}.

\begin{proposition}
  \label{prop:emptyseqcon}
  If $\pat = \emptyseq \con \pat_2$, the following equalities hold:
  \begin{enumerate}
  \item $\typeof{1}{\pat}{C} = \typeof{\emptyw}{\emptyseq}{C /
    L(\pat_2)}$
  \item $\typeof{2n}{\pat}{C} = \typeof{n}{\pat_2}{C}$
  \item $\br{\emptyw}{\pat}{C} = \typeof{1}{\pat}{C} \con \{\cm\} \con
    \typeof{2}{\pat}{C}$
  \item $\br{2n}{\pat}{C} = \br{n}{\pat_2}{C}$
  \end{enumerate}  
\end{proposition}

\begin{proposition}
  \label{prop:sigmacon}
  If $\pat = \sigma \con \pat_2$, the following equalities hold:
  \begin{enumerate}
  \item $\typeof{1}{\pat}{C} = \typeof{\emptyw}{\sigma}{C /
      L(\pat_2)}$ 
  \item $\typeof{2n}{\pat}{C} = \typeof{n}{\pat_2}{ (L(\sigma) \backslash
      C)}$
  \item $\br{\emptyw}{\pat}{C} = \typeof{1}{\pat}{C} \con \{ \cm \}
    \con \typeof{2}{\pat}{C}$ 
  \item $\br{2n}{\pat}{C} = \br{n}{\pat_2}{L(\sigma) \backslash C}$
  \end{enumerate}
\end{proposition}

For the case where $\pat = \pat_1^* \con \pat_2$ the situation is a
bit more involved:

\begin{proposition}
  \label{prop:kleenecon}
  When $\pat = \pat_1^* \con \pat_2$ the following equalities hold:
  \begin{enumerate}
  \item $\typeof{1}{\pat}{C} = T_1$
  \item $\typeof{2n}{\pat}{C} = \typeof{n}{\pat_2}{C_2}$
  \item $\br{\emptyw}{\pat}{C} = I$
  \item $\br{2n}{\pat}{C} = \br{n}{\pat_2}{C_2}$
  \end{enumerate}
  Here, $T_1 = \{ p \in L(\pat_1^*) \mid \exists s \in L(\pat_2): ps
  \in C \wedge c \}$, $C_2 = \{ s \in L(\pat_2) \mid \exists p \in
  L(\pat_1^*): ps \in C \wedge c \}$ and $I = \{ p \cm s \mid ps \in C
  \wedge p \in L(\pat_1^*) \wedge s \in L(\pat_2) \wedge c \}$ with $c
  \equiv \neg(\exists w_3, w_4: w_3 \not = \emptyw \wedge w_3w_4 = s
  \wedge pw_3 \in L(\pat_1^*) \wedge w_4 \in L(\pat_2))$.
\end{proposition}
Of course, this proposition is of little use if we cannot calculate
$T_1$, $C_2$ and $I$. The next lemma gives one possible way of
calculating them and also shows that they are regular if $C$ is. We
denote $\labs \cup \{\cm\}$ by $\labsc$ and let $\pi$ be the
homomorphism from $\labsc$ to $\labs$ with $\pi(\sigma) =
\sigma$ for every $\sigma \in \labs$ and $\pi(\cm) = \emptyw$.
Clearly, if $L$ is a regular language, so is $\ipi(L) = \{\w \in
\labs_{\cm}^* \mid \pi(\w) \in L\}$.

\begin{lemma}
  \label{lemma:calc}
  Using the notation of Proposition~\ref{prop:kleenecon}, and writing 
  $L(\pat_1^*)$ as $L_1$, $L(\pat_2)$ as $L_2$ and $\ipi(L_1) - (L_1
  \con \{ \cm \})$ as $A$, we have:
  \begin{itemize}
  \item $I = \ipi(C) \cap ((L_1 \con \{\cm\} \con L_2) - A \con L_2)$,
  \item $T_1 = I / (\{\cm\} \con L_2)$, and
  \item $C_2 = (L_1 \con \{\cm\}) \backslash I$.
  \end{itemize}
\end{lemma}
\begin{proof}
  By definition, $(L_1 \con \{\cm\} \con  L_2) - A \con L_2$ equals
  \begin{equation*}
    \{\w_1 \cm \w_2 \mid \w_1 \in L_1 \wedge \w_2 \in L_2
    \wedge \neg(\exists v_1, v_2: v_1 v_2 = \w_1 \cm \w_2 \wedge v_1
    \in A \wedge v_2 \in L_2)\} 
  \end{equation*}
  Or, more elaborately,
  \begin{equation*}
    \begin{split}
      \{\w_1 \cm \w_2 \mid &\ \w_1 \in L_1 \wedge\ \w_2 \in
      L_2\ \wedge\ \neg(\exists v_1, v_2: v_1 v_2 = \w_1
      \cm \w_2 \\ & \wedge \pi(v_1) \in L_1 \wedge (\forall p \in
      L_1: v_1 \not = p \cm) \wedge v_2 \in L_2)\}
    \end{split}
  \end{equation*}

  We show that this equals 
  \begin{equation*} 
    \begin{split}
      \{\w_1 \cm \w_2 \mid &\ \w_1 \in L_1 \wedge\ \w_2 \in
      L_2\ \wedge \\ & \neg(\exists \w_3, \w_4: \w_3 \not = \emptyw \wedge
      \w_2 = \w_3 \w_4 \wedge \pi(\w_1 \cm \w_3) \in
      L_1 \wedge \w_4 \in L_2)\}
    \end{split}
  \end{equation*}
  
  We can see this as follows. Suppose $\w_1 \cm \w_2$ is in
  the upper set and suppose that there do exist $\w_3$ and $\w_4$ such
  that $\w_2 = \w_3 \w_4$, $\w_3 \not = \emptyw$, $\pi(\w_1 
  \cm \w_3) \in L_1$ and $\w_4 \in L_2$. Then take $v_1 = \w_1
  \cm  \w_3$ and $v_2 = \w_4$ to see that $\w_1 \cm
  \w_2$ cannot be in the upper set, a contradiction. On the other
  hand, suppose $\w_1 \cm \w_2$ is in the lower set and
  suppose that there do exist $v_1$ and $v_2$ such that $v_1 v_2
  = \w_1 \cm \w_2$, $\pi(v_1) \in L_1$, $\forall p \in L_1:
  v_1 \not = p \cm$ and $v_2 \in L_2$.  Since $v_2 \in L_2$ and
  $L_2$ is a language over $\labs$, $v_2$ cannot contain the
  symbol $\cm$. Since $v_1 v_2 = \w_1 \cm  \w_2$, $v_2$
  must be a suffix of $\w_2$. Hence, we can divide $\w_2$ in $\w_3$
  and $\w_4$ such that $v_1 = \w_1  \cm  \w_3$ and $v_2 = \w_4$.
  Since $v_1 \not = p  \cm$ for any $p$, $\w_3$ must be different
  from $\emptyw$. Moreover, we immediately have $ \pi(\w_1  \cm
   \w_3) = \pi(v_1) \in L_1$ and $\w_4 = v_2 \in L_2$, which gives
  us a contradiction. 
  
  As a consequence, $\ipi(C) \cap ((L_1 \con \{\cm\} \con L_2) - A \con L_2)$
  must equal
  \begin{equation*}
    \begin{split}
      \{ \w_1 \cm  \w_2 \mid & \ \pi(\w_1 \cm \w_2) \in C \wedge \w_1
      \in L_1 \wedge \w_2 \in L_2 \wedge\ \\ 
      & \neg( \exists \w_3, \w_4: \w_3 \not = \emptyw \wedge \w_3 
      \w_4 = \w_2 \wedge \pi(\w_1 \cm \w_3) \in L_1 \wedge
      \w_4 \in L_2) \} \\
    \end{split}
  \end{equation*}
  Since $\w_1$ and $\w_2$ do not contain $\cm$, $\w_1 \w_2 = \pi(\w_1
  \cm \w_2) \in C$. By the same reasoning $\w_1 \w_3 = \pi(\w_1 \cm
  \w_3) \in L_1$. Hence, $\ipi(C) \cap ((L_1 \con \{\cm\} \con L_2) -
  A \con L_2) = I$, as desired.  With $c$ as in
  Proposition~\ref{prop:kleenecon} we obtain the other two desired
  equalities:
  \begin{equation*}
    \begin{split}
      I / (\{\cm\} \con L_2) & = \{ p \mid \exists s \in L_2:
      p \cm s \in I \} 
       = \{p \mid \exists s \in L_2: p s \in C \wedge p \in L_1
      \wedge c\}  = T_1\\
      (L_1 \con \{\cm\}) \backslash I & = \{ s \mid \exists p \in L_1:
      p \cm  s \in I\} 
       = \{s \mid \exists p \in L_1: p s \in C \wedge s \in L_2
  \wedge c \} = C_2
    \end{split}
  \end{equation*}
\qed
\end{proof}
The case $\pat = (\pat_1 \con \pat_2) \con \pat_3$ is handled as follows: 
\begin{proposition}
  \label{prop:concon}
  If $\pat = (\pat_1 \con \pat_2) \con \pat_3$ and $\pat' = \pat_1
  \con (\pat_2 \con \pat_3)$, the following equalities hold:
  \begin{enumerate}
  \item $\typeof{11n}{\pat}{C} = \typeof{1n}{\pat'}{C}$
  \item $\typeof{12n}{\pat}{C} = \typeof{21n}{\pat'}{C}$
  \item $\typeof{2n}{\pat}{C} = \typeof{22n}{\pat'}{C}$
  \item $\br{2n}{\pat}{C} = \br{22n}{\pat'}{C}$
  \item $\br{11n}{\pat}{C} = \br{1n}{\pat'}{C}$
  \item $\br{12n}{\pat}{C} = \br{21n}{\pat'}{C}$
  \item $\br{\emptyw}{\pat}{C} = \{ \w_1 \w_2 \cm \con \w_3 \mid
    \w_1 \cm  \w_2 \cm \w_3 \in J \}$
  \item $\br{1}{\pat}{C} = J / (\{\cm\} \con \labs^*)$
  \item $\typeof{1}{\pat}{C} = \pi(\br{1}{\pat}{C})$
  \end{enumerate}
  Where $J = \{ \w_1  \cm  \w_2  \cm  \w_3 \mid \w_1 \cm \w_2 \w_3 \in
  \br{\emptyw}{\pat'}{C}, \w_2 \cm  \w_3 \in \br{2}{\pat'}{C} \}$
\end{proposition}

We note that $J$ is regular if $\br{\emptyw}{\pat'}{C}$ and
$\br{2}{\pat'}{C}$ are.  Indeed, to recognize a word in $J$ we simply
start the automaton for $\br{\emptyw}{\pat'}{C}$. When we read the
first $\cm$ we also start the automaton for $\br{2}{\pat'}{C}$,
running both automata in parallel, and modify the transition relation
of $\br{\emptyw}{\pat'}{C}$ to allow an extra $\cm$ to be read. We
accept if both automata are in a final state.

Finally, we treat $(\pat_1 \cor \pat_2) \con \pat_3$:
\begin{proposition}
  \label{prop:orcon}
  If $\pat = (\pat_1 \cor \pat_2) \con \pat_3$ and $\pat' = \pat_1
  \con \pat_3 \cor \pat_2 \con \pat_3$ the following equalities
  hold:
  \begin{enumerate}
   \item $\typeof{1}{\pat}{C} = \typeof{11}{\pat'}{C} \cup
     \typeof{21}{\pat'}{C}$
   \item $\typeof{11}{\pat}{C} = \typeof{11}{\pat'}{C}$
   \item $\typeof{12}{\pat}{C} = \typeof{21}{\pat'}{C}$
   \item $\typeof{2}{\pat}{C} = \typeof{12}{\pat'}{C} \cup
     \typeof{12}{\pat'}{C}$
   \item $\br{\emptyw}{\pat}{C} = \br{1}{\pat'}{C} \cup
     \br{2}{\pat'}{C}$
   \item $\br{11n}{\pat}{C} = \br{11n}{\pat'}{C}$
   \item $\br{12n}{\pat}{C} = \br{21n}{\pat'}{C}$
   \item $\br{2n}{\pat}{C} = \br{12n}{\pat'}{C} \cup \br{22n}{\pat'}{C}$
  \end{enumerate}
\end{proposition}

The type inference algorithm announced in
Theorem~\ref{theorem:inferalgo} now works as follows if $C$ is
regular. For the base cases $\emptyseq$, $\sigma$ and $\pat^*$, Lemma
\ref{lemma:toptype} allows us to calculate every type, which must be
regular. For the other cases, the propositions above dictate how to
calculate the types by recursion, using only regular operations on
regular sets. The algorithm can be seen to use the well-founded
$\sqsupset$-ordering on patterns introduced in the proof of Theorem
\ref{theorem:welldef} in its recursion, from which its termination
follows. However, the observant reader will note that the last
proposition relates $(\pat_1 \cor \pat_2) \con \pat_3$ with the
$\sqsupset$-incomparable $\pat_1 \con \pat_3 \cor \pat_2 \con \pat_3$.
Termination still follows if we modify the algorithm to combine the
results of Proposition \ref{prop:orcon} with the ones of Proposition
\ref{prop:or} to calculate the type of $(\pat_1 \cor \pat_2) \con
\pat_3$ by recursion on $\pat_1 \con \pat_3$ and $\pat_2 \con \pat_3$.

Related work \cite{lambdare,cduce,subtyp} claimed sound and complete
type inference algorithms for the matching relation with rule CKleene'
described at the end of the previous section. Using the same
counterexample pattern $\pat = (a \cor a\con b)^* \con (b \cor
\emptyseq)$ and string $ab$, these algorithms must compute
$\typeof{1}{\pat}{\{ab\}} = \{a\}$ and $\typeof{2}{\pat}{\{ab\}} =
\{b\}$.  In contrast, our algorithm correctly computes
$\typeof{1}{\pat}{\{ab\}} = \{ab\}$ and $\typeof{2}{\pat}{\{ab\}} = \{
\emptyw \}$ in accordance with the longest match policy.

\section{Unifying Type Inference and Matching}
\label{sec:unifying}

The process of matching $\w$ by $\pat$ (and computing the resulting
associations) can naively be implemented by evaluating the matching
relation of Fig. \ref{fig:matching} in a syntax-directed manner. This
approach, however, is inefficient. When we want to match $\w$ by
$\pat_1^* \con \pat_2$, we have to create subdivisions of $\w$ into
$\w_1$ and $\w_2$ satisfying the premises of rule CKleene. Since there
are $\width{w}$ possible divisions, and since checking the premises
for a possible division requires us to match $\w_1$ by $\pat_1^*$ and
$\w_2$ by $\pat_2$, every letter of $\w$ is scanned at least
$\width{w}$ times in the worst case scenario, giving 
$\Omega(\width{w}^2)$ time complexity.  We will now show how the type
inference algorithm of the previous section can be used to compile a
pattern $\pat$ into an NFA that will allow us to execute the matching
process in $O(\width{w})$ time. As a bonus, the computed NFA contains
the inferred type of every node in $\pat$. The NFA can be at least
exponentially larger than $\pat$, but this is not abnormal; indeed,
the same happens in ML when the input is only allowed to be
investigated once \cite{patmatch,backtracking}.

A \emph{non-deterministic finite automaton} (NFA) $A$ is a tuple
$(Q_A, I_A, F_A, \delta_A)$ where $Q_A$ is a set of states, $I_A
\subseteq Q_A$ is the set of initial states, $F_A \subseteq Q_A$ is
the set of final states and $\delta: Q_A \times \labs \cup \{ \emptyw
\} \times Q_A$ is the transition relation. An accepting run of $A$ on
$\w = \sigma_1 \dots \sigma_n$ is a sequence $(q_0, k_0), \dots, (q_m,
k_m)$ where $k_0= 0$, $q_0 \in I_A$, $k_m = n$, $q_m \in F_A$ and for
every $i$ either $\delta_A(q_i, \sigma_{k_{i+1}},q_{i+1})$ with
$k_{i+1} = k_i+1$ or $\delta_A(q_i, \emptyw,q_{i+1})$ with $k_{i+1} =
k_i$ and $q_i \not = q_{i+1}$. The language of $A$ is the set of words
for which an accepting run exists and will be denoted by $L(A)$. A NFA
is deterministic (a DFA) if $I_A$ is a singleton set, for every $q$
and $\sigma$ there exists exactly one $q'$ for which
$\delta(q,\sigma,q')$, and $\delta(q,\emptyw, q')$ iff $q = q'$.  If
$S \subseteq Q_A$ and $\tau = (q_1, k_1), \dots, (q_m, k_m)$ is a run
of $A$ on some word $\w$, then $\pos{\tau}{S} = \{ k_l \mid q_l \in S
\}$. If $\pos{\tau}{S} \not = \emptyset$ then $\bound{\tau}{S}$ is the
couple $(i, j)$ for which $i = \min(\pos{\tau}{S})$ and $j =
\max(\pos{\tau}{S})$. It is $(-1,-1)$ otherwise. The subword of $\w =
\sigma_1 \dots \sigma_n$ bounded by $(i,j)$, denoted as $w_{(i,j)}$ is
$\sigma_{i+1} \dots \sigma_j$ if $0 \leq i \leq j \leq n$ and $\ndef$
otherwise.

If $A_1$ and $A_2$ are NFA's, we write $A_1 \con A_2$ for the
automaton recognizing $L(A_1) \con L(A_2)$ obtained by connecting the
final states of $A_1$ to the initial states of $A_2$ by
$\emptyw$-transitions. We write $A_1 \cup A_2$ for the automaton
recognizing $L(A_1) \cup L(A_2)$ obtained by taking the tuple-wise
union of $A_1$ and $A_2$, and we write $A_1 \cap A_2$ for the
automaton recognizing $L(A_1) \cap L(A_2)$, obtained by the product
construction. In particular, $A_1 \cap A_2$ has $Q_{A_1} \times
Q_{A_2}$ as the set of states.  We denote the minimal DFA recognizing
$\{\cm\}$ as $A_{\cm}$ and let $\pi(A)$ be the automaton where we
transform every $\cm$-transition of $A$ into a $\emptyw$-transition.

Algorithm \ref{alg:hyperaut} computes (recursively) the
\emph{hyperautomaton} $H(\pat,C) = (A,f)$ for pattern $\pat$ and
context $C$.  Here $A$ is an NFA and $f$ is a function relating
bindable nodes $n$ of $\pat$ to triples $(Q_n, I_n, F_n)$, where
$Q_n$, $I_n$ and $F_n$ are all subsets of $Q_A$. We use $(Q_1, I_1, F_1)
\times Q_2$ to denote $(Q_1 \times Q_2, I_1 \times Q_2, F_1 \times
Q_2)$ and $(Q_1, I_1, F_1) \cup (Q_2, I_2, F_2)$ to denote $(Q_1 \cup
Q_2, I_1 \cup I_2, F_1 \cup F_2)$. We now state:
\begin{algorithm}[tb]
  \caption{Calculate the hyperautomaton $H(\pat,C)$.}
  \label{alg:hyperaut}

\begin{algorithmic}[1]
\small
\IF{$\pat = \emptyseq$, $\pat = \sigma$ or $\pat = \pat_1^*$}
  \STATE compute a DFA $A$, recognizing $L(\pat) \cap C$
  \STATE return $(A, f)$ with $f(\emptyw) = (Q_{A},I_{A}, F_{A})$
\ELSIF{$\pat = \pat_1 \cor \pat_2$}
   \STATE compute $(A_1,f_1) = H(\pat_1, C)$ and $(A_2,f_2) = H(\pat_2, C
   - L(\pat_1))$
   \STATE let $A = A_1 \cup A_2$
   \STATE {\bf return} $(A, f)$ with $f(\emptyw) = (Q_A,
   I_A, F_A)$, $f(1n) = f_1(n)$ and $f(2n) = f_2(n)$
\ELSIF{$\pat = \pat_1 \con \pat_2$ with $\pat_1 = \emptyseq$ or
  $\pat_1 = \sigma$ }
   \STATE compute $(A_1, f_1) = H(\pat_1, C / L(\pat_2))$ and
   $(A_2, f_2) = H(\pat_2, L(\pat_1) \backslash C)$
   \STATE let $A = A_1 \con A_2$
   \STATE {\bf return} $(A, f)$ with $f(\emptyw) = (Q_A,
   I_A, F_A)$, $f(1n) = f_1(n)$ and $f(2n) = f_2(n)$
\ELSIF{$\pat = \pat_1^* \con \pat_2$}
   \STATE compute DFA's $A_{I}$ and $A_{T_1}$ for $I$ and $T_1$ as
   defined in Proposition~\ref{prop:kleenecon} 
   \STATE compute $(A_2, f_2) = H(\pat_2, C_2)$ with $C_2$ as in
   Proposition~\ref{prop:kleenecon}
   \STATE let $A = \pi((A_{T_1} \con A_{\cm} \con A_2) \cap A_{I})$ 
   \STATE {\bf return} $(A, f)$ where $f(\emptyw) = (Q_A, I_A, F_A)$,
   $f(1) = (Q_{T_1}, I_{T_1}, F_{T_1}) \times Q_{A_{I}}$ and
   $f(2n) = f_2(n) \times Q_{A_{I}}$ 
   \ELSIF{$\pat = (\pat_1 \con \pat_2) \con \pat_3$}
   \STATE compute $(A',f') = H(\pat_1 \con (\pat_2 \con \pat_3), C)$
   and let $ f'(n) = (Q'_n, I'_n, F'_n)$.
   \STATE let $Q_1 = Q'_1 \cup Q'_{21}$, $I_1 = (Q'_1 \cup
   Q'_{21}) \cap I_{A'}$ and $F_1 = \{q \in F'_{21} \mid \exists q'
   \in I'_{22}:$ there is a $\emptyw$-labeled path from $q$ to $q'$
   and there is a path from  $q'$ to some state in $F_{A'}\}$
   \STATE {\bf return} $(A',f)$ where $f(\emptyw) = f'(\emptyw)$,
   $f(2n) = f'(22n)$, $f(11n) = f'(1n)$, $f(12n) = f'(21n)$ and $f(1) =
   (Q_1,I_1, F_1)$
\ELSIF{$\pat = (\pat_1 \cor \pat_2) \con \pat_3$}
   \STATE compute $(A',f') = H(\pat_1 \con \pat_3 \cor \pat_2 \con \pat_3, C)$
   \STATE {\bf return} $(A, f)$ where $f(\emptyw) =
   f'(\emptyw)$, $f(11) = f'(11)$, $f(12) = f'(21)$,
   $f(1) = f'(11) \cup f'(21)$  and $f(2)
   = f'(12) \cup f'(22)$
\ENDIF
\end{algorithmic}
\end{algorithm}
\begin{theorem}
  \label{theorem:hyperaut}
  The hyperautomaton $(A,f)$ computed by Algorithm \ref{alg:hyperaut}
  has the following properties, where $f(n) = (Q_n, I_n, F_n)$:
  \begin{enumerate}
  \item $L(Q_n, I_n, F_n, \delta_A) = \typeof{n}{\pat}{C}$
  \item \label{it:unambig} The automaton $A$ is unambiguous: for every
    $\w$ there is at most one accepting run of $A$ on $\w$.
  \item \label{it:assoc} For every $\w$: $\w \in C$ and
    $\match{\w}{\pat}{V}$ holds iff there exists an accepting run
    $\tau$ of $A$ on $\w$ and $V(n) = \w_{\bound{\tau}{Q_n}}$ for
    every $n \in \bn{\pat}$
  \end{enumerate}
\end{theorem}
The proof is in Appendix \ref{app:proofhyperaut}.  Part
(\ref{it:assoc}) allows us to derive the associations efficiently if
we are given the accepting run $\tau$ on $w = \sigma_1 \dots
\sigma_l$. The problem is that we are not given this run, but need to
compute it. How can we best determine $V(n)$ in that case? We first
compute the sets of states $S_0, \dots, S_l$ where $S_0 = I_A$ and
$S_{i+1} = \hat{\delta}_A(S_i, \sigma_{i+1})$. Here
$\hat{\delta}_A(S, \sigma)$ is defined to be the set of states
that we can reach from a state in $S$ through a path labeled with
$\sigma$. If $S_l \cap F_A \not = \emptyset$ then $\w$ is
matched by $\pat$. We can then compute $S'_0, \dots, S'_{n}$ with
$S'_{n} = S_l \cap F_A$ and $S'_{i-1} = \hat{\delta}^{-1}_A(S'_i,
\sigma_i) \cap S_{i-1}$ where $\hat{\delta}^{-1}_A(S', \sigma) = S$ if
$\hat{\delta}_A(S, \sigma) = S'$. Intuitively, $S'_i$ contains those
states that can be reached from a start state with a path spelling
$\sigma_1 \dots \sigma_i$ and from which we can reach a state in $F_A$
through a path spelling $\sigma_{i+1} \dots \sigma_l$. Note that we
must be able to order the states of $S'_i$ into a sequence $q_1,
\dots, q_k$ such that $\delta(q_j, \emptyw, q_{j+1})$ or we can
construct multiple accepting runs, contradicting (\ref{it:unambig}).
In particular, if $q^i_1, \dots, q^i_{k_i}$ is the ordering for
$S'_i$, then $\tau = (q^0_1,0), \dots, (q^0_{k_1},0), (q^1_1,1),
\dots, (q^1_{k_1},1), \dots, (q^l_1,l), \dots (q^l_{k_l},l)$ must be
the accepting run. Clearly, $\bound{\tau}{Q_n} = (i, j)$ iff $S'_i
\cap Q_n \not = \emptyset$, $S'_j \cap Q_n \not = \emptyset$ and there
is no $m < i$ or $m > j$ for which $S'_m \cap Q_n \not = \emptyset$.
So, it suffices to search the first and last $i$ for which $S'_{i}
\cap Q_n \not = \emptyset$ to obtain $\bound{\tau}{Q_n}$, which gives
$V(n)$ by (\ref{it:assoc}).  Since $A$ is constant, all the
calculations can be done in $O(\width{\w})$ time, which gives us an
efficient matching implementation.

\section{Future Work}
\label{sec:discussion}
We note that Perl has two ways of disambiguating the Kleene closure in
a concatenation. One of them is the longest match strategy presented
here, making the $*$-operator ``greedy''. Another version of the
$*$-operator, denoted $*?$ in Perl, has a \emph{shortest match}
semantics. That is, it will match the smallest prefix that still
allows the rest of the word to be matched. The matching relation and
the type inference algorithm can be expanded in a straight-forward
manner to include this operator.

Although there has been extensive research on the regular type
inference problem, there has not yet been a formal investigation of
the inherent time complexity bounds. While we conjecture that our
algorithm executes in 2EXPTIME, a formal proof still has to be given.
As we noted in the previous section, the construction of $H(\pat, C)$
can involve an exponential blowup. A tight upper bound
(single-exponential, double-exponential or more) still has to be
found.

The true power of regular expression pattern matching emerges when we
introduce tree patterns matching unranked hedges, in the context of
XML.  We are currently trying to expand the matching relation and type
inference algorithm to this end.  It will be interesting to see if we
can also unify type inference and pattern matching in this setting.

\paragraph{Acknowledgments}
I thank Jan Van den Bussche, Dirk Leinders, Wim Martens and Frank
Neven for inspiring discussions and for their comments on a draft
version of this paper.

\bibliographystyle{splncs}
\bibliography{biblio}

\newpage

\appendix

\section{Proofs of claims in Sec. \ref{sec:typeinfer}}
\label{app:proofinfer}
For completeness' sake we present the proofs of the various
propositions in Sec.~\ref{sec:typeinfer}.

\subsection{Proof of proposition \ref{prop:or}}
\begin{proof}
  If $\w \in C$ with $\match{\w}{\pat}{V}$ then there are two possible
  matching derivations. The first one has the form
  \begin{mathpar}
    \small
    \inferrule*[Right=Or1]
    { 
      \inferrule*{\dots}{  
      \match{\w}{\pat_1}{V}}
    }
    {\match{\w}{\pat_1\ \cor\ \pat_2}{V\ \cor\ \pat_2}}
  \end{mathpar}
  From which we immediately obtain $\match{\w}{\pat_1}{V}$. Now
  $\typeof{1n}{\pat}{C} \subseteq \typeof{n}{\pat}{C} $ and
  $\br{1n}{\pat}{C} \subseteq \br{n}{\pat_1}{C}$ since $(V\ \cor\ 
  \pat_2)(1n) = V(n)$ and since $V(1n) = \ndef$ in the other
  derivation. On the other hand, if $\match{\w}{\pat_1}{V}$ then we
  can create the matching derivation above, which means
  $\typeof{n}{\pat_1}{C} \subseteq \typeof{1n}{\pat}{C}$ and
  $\br{n}{\pat_1}{C} \subseteq \br{1n}{\pat}{C}$.

  The second matching derivation looks like
  \begin{mathpar}
    \inferrule*[Right=Or2]
    { 
        \inferrule*{\dots}{\match{\w}{\pat_2}{V}} \\ \w \not \in \pat_1
    }
    {\match{\w}{\pat_1\ \cor\ \pat_2}{\pat_1\ \cor\ V}}
  \end{mathpar}
  So, $\w \in C - L(\pat_1)$ and $\match{\w}{\pat_2}{V}$. Because
  $(\pat_1\ \cor\ V)(2n) = V(n)$, we have $\typeof{2n}{\pat}{C}
  \subseteq \typeof{n}{\pat}{C - L(\pat_1)}$ and $\br{2n}{\pat}{C}
  \subseteq \br{n}{\pat_2}{C - L(\pat_1)}$. On the other hand, if $\w
  \in C - L(\pat_1)$ and $\match{\w}{\pat_2}{V}$, we can create the
  matching derivation above to obtain $\match{\w}{\pat}{\pat_1\ \cor\ 
    V}$, so $\typeof{n}{\pat_2}{C - L(\pat_1)} \subseteq
  \typeof{2n}{\pat}{C}$ and $\br{n}{\pat_2}{C - L(\pat_1)} \subseteq
  \br{2n}{\pat}{C - L(\pat_1)}$.
\end{proof}

\subsection{Proof of proposition \ref{prop:sigmacon}}
\begin{proof}
  Define $C_1 = C / L(\pat_2)$, $C_2 = L(\sigma) \backslash
  C$. If $\w \in C$ with $\match{\w}{\pat}{V}$ then the top of the
  matching derivation must looks like
  \begin{mathpar}
    \small
    \inferrule*[Right=CElem]
    { 
        \inferrule*[Right=CLab]{
          \inferrule*{\dots}{\match{\sigma}{\sigma}{V_1}} \\ 
          \inferrule*{\dots}{\match{\w_2}{\pat_2}{V_2}}
        }
        {
          \match{(\sigma, \w_2)}{\pat}{(V_1, V_2)}
        }
    }
    {\match{\w = \sigma \w_2}{\pat}{V_1 \con V_2}}
  \end{mathpar}
  Since $\sigma \in C_1$ and $\match{\sigma}{\sigma}{V_1}$,
  $\typeof{1}{\pat}{C} \subseteq \typeof{\emptyw}{\pat_1}{C_1}$. If
  $w_1 \in \typeof{\emptyw}{\sigma}{C_1}$, then $w_1 \in L(\sigma)$
  and there must exist some $\w_2 \in L(\pat_2)$ for which $\w_1 \w_2
  \in C$. Then we can reconstruct the matching derivation above using
  theorem \ref{theorem:welldef}.\ref{it:correct} and so equality (1)
  holds.
  
  Since $\w_2 \in C_2$ and $\match{\w_2}{\pat_2}{V_2}$,
  $\typeof{2n}{\pat}{C} \subseteq \typeof{n}{\pat_2}{C_2}$ and
  $\br{2n}{\pat}{C} \subseteq \br{n}{\pat_2}{C_2}$ hold. On the other
  hand, if $\w_2 \in C_2$ then by definition $\sigma \w_2 \in C$.
  If $\match{\w_2}{\pat_2}{V_2}$, then we can create the matching
  derivation above, from which we may conclude that
  $\typeof{n}{\pat_2}{C_2} \subseteq \typeof{2n}{\pat}{C}$,
  $\br{n}{\pat_2}{C_2} \subseteq \br{2n}{\pat}{C}$ and
  $\typeof{1}{\pat}{C} \con \{\cm\} \con \typeof{2}{\pat}{C} \subseteq
  \br{\emptyw}{\pat}{C}$. The inclusion 
\[\br{\emptyw}{\pat}{C} \subseteq \typeof{1}{\pat}{C} \con \{\cm\}
  \con \typeof{2}{\pat}{C}\] 
trivially holds.
\end{proof}

Proposition \ref{prop:emptyseqcon} can be proven in a similar way.

\subsection{Proof of proposition \ref{prop:kleenecon}}
\begin{proof}
  Let $\w \in C$ and suppose $\match{\w}{\pat}{V}$, then the top of
  the matching derivation must look like
  \begin{mathpar}
    \small
    \inferrule*[Right=CElem]
    {
      \inferrule*[Right=CKLeene]
      { \inferrule*{\dots}{\match{\w_1}{\pat_1^*}{V_1}}
        \\ \inferrule*{\dots}{\match{\w_2}{\pat_2}{V_2}} \\\\
        \neg(\exists w_3 \not = \emptyw, \w_4: \w_2 = \w_3 \w_4
        \wedge \w_1 \w_3 \in \pat_1^* \wedge \w_4 \in \pat_2) 
      }        
      {\match{(\w_1, \w_2)}{\pat}{(V_1,V_2)}}
    }
    {
      \match{\w = \w_1 \w_2}{\pat}{V_1 \con V_2}
    }
  \end{mathpar}
  The first equality then holds by this observation and application of
  Theorem \ref{theorem:welldef}.\ref{it:correct}.  Since $\w_2 \in
  C_2$, $\typeof{2n}{\pat}{C} \subseteq \typeof{n}{\pat_2}{C_2}$ and
  $\br{2n}{\pat}{C} \subseteq \br{n}{\pat}{C_2}$. On the other hand,
  if $\w_2 \in C_2$, we know that $\w_2 \in L(\pat_2)$, and there must
  exist some $\w_1 \in L(\pat_1)$ with $\w_1 \w_2 \in C$ for which
  condition $c$ holds. By application of Theorem
  \ref{theorem:welldef}.\ref{it:correct} we can reconstruct the
  matching derivation shown above. As such $\typeof{n}{\pat_2}{C_2}
  \subseteq \typeof{2n}{\pat}{C}$ and $\br{n}{\pat_2}{C_2} \subseteq
  \br{2n}{\pat_2}{C}$.
  
  If $\w_1 \cm  \w_2 \in \br{\emptyw}{\pat}{C}$ then, by
  definition, there exists some $\w \in C$ such that
  $\match{\w}{\pat}{V}$, $V(1) = \w_1$ and $V(2) = \w_2$. Then, by the
  derivation above and Theorem \ref{theorem:welldef}.\ref{it:correct}
  we know that $\w = \w_1  \w_2$, $\w_1 \in L(\pat_1^*)$ and $\w_2
  \in L(\pat_2)$.  Then $\w_1  \cm  \w_2 \in I$ by applying
  the same theorem to the third premise of CKleene. On the other hand,
  if $\w_1 \cm  \w_2 \in I$, we can use the theorem again to
  reconstruct the matching derivation above, showing that $I \subseteq
  \br{\emptyw}{\pat}{C}$.
  \qed
\end{proof}

\subsection{Proof of proposition \ref{prop:concon}}
We will use the following lemma:
\begin{lemma}
  \label{lemma:consecmatch}
  If $\match{(\w_1, \w_2)}{\pat}{(V_1,V_2)}$ then
  $\match{\w_2}{\pat}{V_2}$
\end{lemma}
\begin{proof}
  The proof goes by induction on the matching derivation
  $\match{(\w_1, \w_2)}{\pat}{V_1,V_2}$ with a case analysis on the
  last rule used. In all the cases, the result either follows
  immediately from the premise of the last rule used, or follows
  immediately from the induction hypothesis.  \qed
\end{proof}

\begin{proof}
  If $w \in C$ with $\match{\w}{\pat}{V}$, then the matching
  derivation must look like
  \begin{mathpar}
    \small
    \inferrule*[Right=CElem]
    { 
      \inferrule*[Right=CCon]
      { 
        \inferrule* 
        { 
          \dots
        }
        {\match{(\w_1, \w_2 w_3)}{\pat_1 \con (\pat_2 \con
            \pat_3)}{(V_1,W)}
        } \\
       \inferrule*
        {\dots}
        {\match{(\w_2, \w_3)}{\pat_2 \con \pat_3}{(V_2, V_3)}}
      }
      {
        \match{(\w_1 \w_2, \w_3)}{(\pat_1 \con \pat_2) \con
            \pat_3}{(V_1 \con V_2, V_3)}
        }
    }
    {\match{\w = \w_1 \w_2 \w_3}{\pat_1\con \pat_2}{(V_1 \con
    V_2) \con V_3}}
  \end{mathpar}
  Using CElem on the second premise of CCon, we obtain
  $\match{\w_2\w_3}{\pat_2 \con \pat_3}{V_2 \con V_3}$. By lemma
  \ref{lemma:consecmatch} $\match{\w_2\w_3}{\pat_2 \con \pat_3}{W}$,
  which means $W = V_2 \con V_3$ by Theorem \ref{theorem:welldef}.  By
  using CElem on the first premise of CCon, $\match{\w}{\pat'}{V_1
    \con (V_2 \con V_3)}$.  If $\match{\w}{\pat'}{V}$, the derivation
  must look like
  \begin{mathpar}
    \small
    \inferrule*[Right=CElem]
    {
      \inferrule*
      { \dots
      }
      {\match{(\w_1, \w_2 w_3)}{\pat_1 \con (\pat_2 \con
          \pat_3)}{(V_1,W)}
      }
    }
    {
      \match{w = \w_1 \w_2 \w_3}{\pat_1 \con (\pat_2 \con
          \pat_3)}{V_1 \con W}
    }
  \end{mathpar}
  We know by lemma \ref{lemma:consecmatch} that
  $\match{\w_2\w_3}{\pat_2 \con \pat_3}{W}$. By the premise of CElem,
  that can only happen if $\match{(\w_2, \w_3)}{\pat_2 \con
    \pat_3}{V_2, V_3}$ for some $V_2$ and $V_3$.  But then, we can use
  CCon to obtain $\match{\w}{\pat}{(V_1 \con V_2) \con V_3}$.
  
  From these observations, equalities (1) until (6) follow
  immediately.
  
  Next, we we will show that $J= \{ \w_1 \cm \w_2 \cm \w_3 \mid \w_1
  \w_2 \w_3 \in C, \match{\w_1 \w_2 \w_3}{\pat}{V}, V(1) = \w_1 \w_2,
  V(11) = \w_1, V(12) = \w_2, V(2) = \w_3\}$. Indeed, suppose $\w_1
  \cm \w_2 \w_3 \in \br{\emptyw}{\pat'}{C}$ and $\w_2 \cm \w_3 \in
  \br{2}{\pat'}{C}$. This means there must exist some $\w' \in C$ with
  $\match{\w'}{\pat'}{V}$. As explained above, $V = V_1 \con (V_2 \con
  V_3)$. Since $V(1) = \w_1$ and $V(2) = \w_2 \w_3$, $\w' = \w_1 \w_2
  \w_3$. Since $\w_2 \cm \w_3 \in \br{2}{\pat'}{C}$, there must exist
  some $\w'' \in C$ with $\match{\w''}{\pat'}{V'}$, $V'(21) = \w_2$
  and $V'(22) = \w_3$. By the observations about the matching
  derivation of $\pat'$ made above, $\w'' = v \w_2 \w_3$ and $V' =
  V'_1 \con (V'_2 \con V'_3)$. Then $\match{\w_2 \w_3}{\pat_2 \con
    \pat_3}{V'_2 \con V'_3}$ and by lemma \ref{lemma:consecmatch}
  $\match{\w_2 \w_3}{\pat_2 \con \pat_3}{V_2 \con V_3}$, so $V_2 \con
  V_3 = V'_2 \con V'_3$ by Theorem \ref{theorem:welldef}. Thus, $V(21)
  = \w_2$ and $V(22) = \w_3$. As such, $J = \{ \w_1 \cm \w_2 \cm \w_3
  \mid \w_1 \w_2 \w_3 \in C, \match{\w_1 \w_2 \w_3}{\pat'}{V}, V(1) =
  \w_1, V(21) = \w_2, V(22) = \w_3\}$.  By the observations made
  above, this equals $\{\w_1 \cm \w_2 \cm \w_3 \mid \w_1 \w_2 \w_3 \in
  C, \match{\w_1 \w_2 \w_3}{\pat}{V}, V(1) = \w_1 \w_2, V(11) = \w_1,
  V(12) = \w_2, V(2) = \w_3\}$.
  
  The $\supseteq$ inclusion of equalities (7) and (8) then follows
  immediately. Now suppose $\w \in C$, $\match{\w}{\pat}{V}$ and $V(1)
  = \w_1 \w_2, V(2) = \w_3$. As we can see in the matching derivation
  above, $\w = \w_1 \w_2 \w_3$, so $\w_1 \cm \w_2 \cm \w_3 \in J$,
  from which we may conclude $\br{\emptyw}{\pat}{C} \subseteq \{ \w_1
  \w_2 \cm \w_3 \mid \w_1 \cm \w_2 \cm \w_3 \in J \}$. If $\w \in C$,
  $\match{\w}{\pat}{V}$ with $V(11) = \w_1$ and $V(12) = \w_2$, then
  $V(1) = \w_1 \w_2$ and there must exist some $\w_3$ such that $\w_1
  \w_2 \w_3 \in C$ with $V(2) = \w_3$ (see the derivation above). So,
  $\br{1}{\pat}{C} \subseteq J / (\{\cm\} \con \labs^*)$. The
  last equality follows directly from equality (8).
\end{proof}

\subsection{Proof of Proposition \ref{prop:orcon}}

\begin{proof}
  If $\w \in C$ and $\match{\w}{\pat}{V}$ then there are two
  possibilities for the top of the matching derivation:
  \begin{enumerate}
  \item Rule COr1 is used:
    \begin{mathpar}
      \small
      \inferrule*[Right=CElem]
      { 
        \inferrule*[Right=COr1]
        { 
          \inferrule* 
          { 
            \dots
          }
          {\match{(\w_1, \w_2)}{\pat_1 \con \pat_3}{(V_1,V_2)}
          }
        }
        {
          \match{(\w_1, \w_2)}{(\pat_1\ \cor\ \pat_2) \con
            \pat_3}{(V_1\ \cor\ \pat_2, V_2)}
        }
      }
      {\match{\w = \w_1 \w_2}{(\pat_1\ \cor\ \pat_2) \con
          \pat_3}{(V_1\ \cor \pat_2) \con V_2}}
    \end{mathpar}
    
    So we can create the following matching derivation, proving
    $\match{\w}{\pat'}{(V_1 \con V_2)\ \cor\ (\pat_2 \con \pat_3)}$
    \begin{mathpar}
      \small
      \inferrule*[Right=Or1]
      { 
        \inferrule*[Right=CElem]
        { 
          \inferrule* 
          { 
            \dots
          }
          {\match{(\w_1, \w_2)}{\pat_1 \con \pat_3)}{(V_1,V_2)}
          }
        }
        {
          \match{\w_1 \w_2)}{\pat_1 \con \pat_3}{V_1 \con V_2}
        }
      }
      {\match{\w = \w_1 \con \w_2}{(\pat_1 \con \pat_3)\ \cor\ 
          (\pat_2 \con \pat_3)}{(V_1 \con V_2)\ \cor\ (\pat_2 \con \pat_3)}}
    \end{mathpar}
    On the other hand, suppose $\match{\w}{\pat'}{V'}$ for some $V'$,
    then the matching derivation could be of the same form as above.
    If it is we can easily construct the first matching derivation
    given here, so $\match{\w}{\pat}{(V_1\ \cor \pat_2) \con V_2}$.
    
  \item Rule $COr2$ is used:
    \begin{mathpar}
      \small
      \inferrule*[Right=CElem]
      { 
        \inferrule*[Right=CCor2]
        { 
          \inferrule* 
          { 
            \dots
          }
          {\match{(\w_1, \w_2)}{\pat_2 \con \pat_3}{(V_1,V_2)}
          }\\
          \w_1 \w_2 \not \in \pat_1 \con \pat_3
        }
        {
          \match{(\w_1, \w_2)}{(\pat_1\ \cor\ \pat_2) \con
            \pat_3}{(\pat_1\ \cor\ V_1, V_2)}
        }
      }
      {\match{\w = \w_1  \w_2}{(\pat_1\ \cor\ \pat_2) \con
          \pat_3}{(\pat_1\ \cor V_1) \con V_2}}
    \end{mathpar}
    Then we can make the following matching derivation to show that
    $\match{\w}{\pat'}{(\pat_1 \con \pat_3)\ \cor\ (V_1 \con V_2)}$
    \begin{mathpar}
      \small
      \inferrule*[Right=Or2]
      { 
        \inferrule*[Right=CElem]
        { 
          \inferrule* 
          { 
            \dots
          }
          {\match{(\w_1, \w_2)}{\pat_2 \con \pat_3}{(V_1,V_2)}
          }
        }
        {
          \match{\w_1 \con \w_2)}{\pat_2 \con \pat_3}{V_1 \con V_2}
        }\\
        \w_1 \w_2 \not \in \pat_1 \con \pat_3
      }
      {\match{\w = \w_1 \con \w_2}{(\pat_1 \con \pat_3)\ \cor\ 
          (\pat_2 \con \pat_3)}{(\pat_1 \con \pat_3)\ \cor\ (V_1 \con V_2)}}
    \end{mathpar}
    On the other hand, if $\match{\w}{\pat'}{V'}$ and we do not use
    Or1 at the top (which is suggested in the previous case), the
    matching derivation must look like the one given above. But then
    we can easily create the matching derivation at the beginning of
    this case to show that $\match{\w}{\pat}{(\pat_1\ \cor\ V_1) \con
      V_2}$.
  \end{enumerate}
  
  All equalities follow from these observations.
\end{proof}

\section{Proof of theorem \ref{theorem:hyperaut}}
\label{app:proofhyperaut}
We will prove the following, stronger theorem:

\begin{theorem}
  The hyperautomaton $(A,f)$ computed by Algorithm \ref{alg:hyperaut}
  has the following properties, where $f(n) = (Q_n, I_n, F_n)$:
  \begin{enumerate}
  \item \label{it:ptype} $L(Q_n, I_n, F_n, \delta_A) = \typeof{n}{\pat}{C}$.
  \item \label{it:punambig} The automaton A is unambiguous: for every
    $\w$ there is at most one accepting run of $A$ on $\w$.
  \item \label{it:passoc}For every $\w$: $\w \in C$ and
    $\match{\w}{\pat}{V}$ holds iff there exists an accepting run
    $\tau$ of $A$ on $\w$ and $V(n) = \w_{\bound{\tau}{Q_n}}$ for
    every $n \in \bn{\pat}$.
  \item \label{it:pbound}If $n \in \bn{\pat}$, $\pat(n) = \con$,
    $\tau$ is an accepting run of $A$, $\bound{\tau}{Q_{n1}} = (i_1,
    j_1)$ and $\bound{\tau}{Q_{n2}} = (i_2, j_2)$ (both different from
    $(-1, -1)$), then $j_1 = i_2$ and we can find $q \in F_{n1}$ and
    $q' \in I_{n2}$ such that $(q,j_1), \dots, (q', i_2)$ appears in
    $\tau$.
  \end{enumerate}  
\end{theorem}
\begin{proof}
  The proof goes by well-founded induction on $(\mathcal{P},
  \sqsupset)$, where the $\sqsupset$ ordering on patterns was
  introduced in the proof of Theorem \ref{theorem:welldef}.
  
  \paragraph{Base case}
  We treat $\pat = \emptyseq$, $\sigma$ or $\pat_1^*$ together.
  Property (\ref{it:ptype}) follows directly from the construction and
  Lemma \ref{lemma:toptype}. Property (\ref{it:punambig}) holds
  trivially since the computed automaton is a DFA. Property
  (\ref{it:passoc}) holds since $\lambda$ is the only bindable
  node. Since there is no bindable node labeled with a concatenation, 
  (\ref{it:pbound}) trivially holds.
  
  \paragraph{$\pat = \pat_1 \cor \pat_2$}
  If $\pat = \pat_1 \cor \pat_2$, let $(A_1,f_1) = H(\pat_1,C)$ and
  $(A_2,f_2) = H(\pat_2,C - L(\pat_1))$. For these hyperautomata, the
  theorem holds by induction. Property (\ref{it:ptype}) follows
  directly from the induction hypothesis and Proposition \ref{prop:or}
  if $n \not = \emptyw$. To show that the property holds for $n =
  \emptyw$, we use lemma \ref{lemma:toptype} and the induction
  hypothesis: $\typeof{\emptyw}{\pat_1}{C} \cup
  \typeof{\emptyw}{\pat_2}{C - L(\pat_1)} = (L(\pat_1) \cap C) \cup
  (L(\pat_2) \cap C - L(\pat_1)) = (L(\pat_1) \cup L(\pat_2)) \cap C =
  \typeof{\emptyw}{\pat}{C}$. Suppose that $A$ is ambiguous, i.e.
  that there are two accepting runs of $A$ on $\w$.  Then one has to
  be an accepting run of $A_1$ on $\w$ and the other an accepting run
  of $A_2$ on $\w$ (otherwise, $A_1$, or $A_2$ would be ambiguous,
  which is impossible by the induction hypothesis). But then, $\w \in
  \typeof{\emptyw}{\pat_1}{C} = L(\pat_1) \cap C$ and $\w \in
  \typeof{\emptyw}{\pat_2}{C - L(\pat_1)} = L(\pat_2) \cap (C -
  L(\pat_1))$, which means $\w \in L(\pat_1)$ and $\w \not \in
  L(\pat_1)$, contradiction. In the proof of Proposition \ref{prop:or}
  we remarked that $\match{\w}{\pat}{V}$ iff either
  $\match{\w}{\pat_1}{V_1}$ with $V = V_1 \cor \pat_2$ or
  $\match{\w}{\pat_2}{V_2}$ with $V = \pat_1 \cor V_2$ and $\w \not
  \in \pat_1$. Property (\ref{it:passoc})then follows directly from
  the induction hypothesis and the fact that the accepting run must
  either be an accepting run of $A_1$ or of $A_2$.  Property
  (\ref{it:pbound}) follows directly from the induction hypothesis.
  
  \paragraph{$\pat = \sigma \con \pat_2$}
  Let $(A_1,f_1) = H(\sigma, C / L(\pat_2))$ and $(A_2, f_2) =
  H(\pat_2, L(\sigma) \backslash C)$. Property (\ref{it:ptype}) follows
  from the induction hypothesis and Proposition \ref{prop:sigmacon} if
  $n \not = \emptyw$. It also holds for $n = \emptyw$ since
  $\typeof{\emptyw}{\pat}{C} = \pi(\br{\emptyw}{\pat}{C})$ and this
  equals $\typeof{\emptyw}{\sigma}{C / L(\pat_2)} \con
  \typeof{\emptyw}{\pat_2}{L(\sigma} \backslash C)$ by Proposition
  (\ref{prop:sigmacon}).
  
  Note that we can split any accepting run $(q_0, k_0), \dots, (q_m,
  k_m)$ of $A$ on $\sigma \w$ into an accepting run $(q_0, k_0),
  \dots, (q_l, k_l)$ of $A_1$ on $\sigma$ and $(q_{l+1}, k_{l+1}),
  \dots (q_m, k_m)$ of $A_2$ on $\w$. If there are multiple such runs
  of $A$, we would get different runs of $A_1$ or $A_2$, which is
  impossible since they are unambiguous, so Property
  (\ref{it:punambig}) holds.
  
  From the proof of Proposition \ref{prop:sigmacon} we know that
  $\match{\sigma \w}{\pat}{V}$ iff $\match{\sigma}{\sigma}{V_1}$ and
  $\match{\w}{\pat_2}{V_2}$ with $V = V_1 \con V_2$. Suppose $\sigma
  \w \in C$ and $\match{\sigma \w}{\pat}{V}$. By the induction
  hypothesis, we have $\tau_1$ of $A_1$ on $\sigma$ with $V_1(\emptyw)
  = \bound{{\tau_1}}{{Q_1}}$ and $\tau_2$ of $A_2$ on $\w$ with
  $V_2(n) = \bound{{\tau_2}}{Q_{2n}}$. Since $\tau_1$ cannot contain
  states of $A_2$ and $\tau_2$ cannot contain states of $\tau_1$, and
  $\tau_1, \tau_2$ is the accepting run of $A$, the ``if'' part of
  (\ref{it:passoc}) follows.  On the other hand, we know that we can
  split any accepting run $\tau$ of $A$ into an accepting run $\tau_1$
  of $A_1$ (where no states of $A_2$ can occur) and an accepting run
  $\tau_2$ of $A_2$ (where not states of $A_1$ can occur). By the
  induction hypothesis, $\match{\sigma}{\sigma}{V_1}$ with
  $V_1(\emptyw) = \sigma$ and $\match{\w}{\pat_2}{V_2}$ with $V_2(n) =
  \w_{\bound{{\tau_2}}{Q_{2n}}} = (\sigma \w)_{\bound{\tau}{Q_{2n}}}$,
  from which the ``only if'' part follows. Property (\ref{it:pbound})
  is clear for $n = \emptyw$ from the remarks made above about an
  accepting run of $A$ and it follows from the induction hypothesis
  otherwise.

  \paragraph{$\pat = \emptyseq \con \pat_2$}
  We note that $L(\emptyseq) \backslash C = C$. The proof is then
  analogous to the previous case.
  
  \paragraph{$\pat = \pat_1^* \con \pat_2$}
  Let $A_I$ be a deterministic automaton for $I$, $A_{T_1}$ be a
  deterministic automaton for $T_1$, $(A_2,f_2) = H(\pat_2, C_2)$ and
  $A = \pi((A_{T_1} \con A_{\cm} \con A_2) \cap A_I)$.  For $n \not =
  \emptyw$, Property (\ref{it:ptype}) follows from Proposition
  \ref{prop:kleenecon} and the fact that we do not lose information
  about the subautomata by taking the intersection and performing
  $\pi$. Moreover, $I \subseteq \typeof{1}{\pat}{C} \con \{ \cm \}
  \con \typeof{2}{\pat}{C}$, so $L(((A_{T_1} \con A_{\cm} \con A_2)
  \cap A_I) = I$. The result then follows for $n = \emptyw$ since
  $\typeof{\emptyw}{\pat}{C} = \pi(\br{\emptyw}{\pat}{C}) = \pi(I)$.
  
  To prove unambiguity, we first note the following. Let $B$ be the
  automaton obtained by computing $(A_{T_1} \con A_{\cm} \con A_2)
  \cap A_I$. We assume that the states of $B$ are of the form $(q,s)$
  with $q \in Q_{A_{T_1} \con A_{\cm} \con A_2}$ and $s \in Q_{A_I}$.
  If $\w \in L(A)$, then there must exist some $\w' \in L(B)$ with
  $\pi(\w) = \w'$. Thus, there are $\w_1 \in L(A_{T_1})$ and $\w_2 \in
  L(A_2)$ such that $\w = \w_1 \w_2$ and $\w' = \w_1 \cm \w_2 \in I$.
  By definition of $I$, $\w_1$ and $\w_2$ are unique.  Let $\tau =
  ((q_0, s_0), k_0), \dots, ((q_m, s_m), k_m)$ be an accepting run of
  $A$ on $\w$. Since $A = \pi(B)$ and since $A_{\cm}$ is minimal, we
  can find exactly one $i$ for which $q_i \in I_{A_{\cm}}$, $q_{i+1}
  \in F_{A_{\cm}}$ and $k_i = k_{i+1} = \width{\w_1}$. Necessarily,
  $q_{i-1} \in F_{T_2}$ and $q_{i+1} \in I_{A_2}$.
  
  Suppose we have two accepting runs $\tau_1 = ((q_0, s_0), k_0),
  \dots, (q_m, s_m), k_m)$ and $\tau_2 = ((q'_0, s'_0), k'_0), \dots,
  ((q'_l, s'_l), k'_l)$ of $A$ on $\w = \sigma_1 \dots \sigma_n$. Then
  we can find $i_1$ and $i_2$ for $\tau_1$ respectively $\tau_2$ as
  described above. By definition of $I$, $k_{i_1} = k'_{\i_2}$. Since
  both $(q_0,k_0), \dots, (q_{i_1 - 1}, k_{i_1 -1})$ and $(q'_0,
  k'_0), \dots, (q'_{i_2 -1}, k_{i_2 -1})$ are accepting runs of
  $A_{T_1}$ on $\sigma_1 \dots \sigma_{k_{i_1}}$, and since $A_{T_1}$
  and $A_I$ are deterministic, $((q_0, s_0), k_0), \dots, ((q_{i_1
    -1}, s_{i_1 -1}), k_{i_1 -1}) = ((q'_0, s'_0), k_0), \dots,
  ((q'_{i_2 -1}, s_{i_2 -1}), k_{i_2 -1})$.  Since the automaton
  $A_{\cm}$ is deterministic, $((q_{i_1},s_{i_1}) , k_{i_1}),
  ((q_{i_1+1},s_{i_1+1}), k_{i_1 + 1}) = ((q'_{i_2},
  s'_{i_2}),k'_{i_2}), ((q'_{i_2+1},s'_{i_2+1}), k'_{i_2 + 1})$.
  Since $A_2$ is unambiguous by induction, the accepting
  runs $(q_{i_1+2},k_{i_1+2}), \dots, (q_m, k_m)$ and $ (q'_{i_2 +2},
  k'_{i_2 + 2}), \dots, (q'_l, k'_l)$ of $A_2$ on $\w_2$ must be
  equal. So, $\tau_1 = \tau_2$, from which Property
  (\ref{it:punambig}) follows.
  
  Suppose $f_2(n) = (Q'_n, I'_n, F'_n)$. By the remarks about the
  matching derivations of $\pat_1^* \con \pat_2$ made in the proof of
  Proposition \ref{prop:kleenecon}, we know that $\match{\w}{\pat}{V}$
  iff $w = \w_1 \w_2$ with $\w_1 \cm \w_2 \in I$,
  $\match{\w_1}{\pat_1^*}{V_1}$ with $V_1(\emptyw) = \w_1$ and
  $\match{\w_2}{\pat_2}{V_2}$ where $V = V_1 \con V_2$. Since $\w_1
  \cm \w_2 \in I$, $\w_2 \in C_2$. Suppose $\w \in C$ and
  $\match{\w}{\pat}{V}$ By the induction hypothesis of Property
  (\ref{it:passoc}), there exists a run $\tau_2$ of $A_2$ on $\w_2$
  with $V_2(n) = {\w_2}_{\bound{{\tau_2}}{Q'_n}}$. By definition of
  $I$ and $T_1$, there exists a run $\tau_1$ of $A_{T_1}$ on $\w_1$
  with ${\w_1}_{\bound{{\tau_1}}{Q_{T_1}}} = \w_1$. Then we can find
  $q \in I_{A_{\cm}}$ and $q' \in F_{A_{\cm}}$ such that $\tau' =
  \tau_1,(q,\width{\w_1}), (q', \width{\w_1}+1), \tau'_2$ is an
  accepting run of $A_1 \con A_{\cm} \con A_2$ on $\w_1 \cm \w_2$
  (here, $\tau'_2$ is obtained from $\tau_2$ by adding $\width{\w_1} +
  1$ to the indexes). Since we know that $\w_1 \cm \w_2 \in I$, there
  is a run $\tau''$ of $A_I$ on it. By combining the $\tau'$ and
  $\tau''$, we can get an accepting run $\tau$ of $A$ on $\w_1 \w_2$.
  It is easy to see that $\w_{\bound{\tau}{Q_{1}}} = \w_1$ and
  $\w_{\bound{\tau}{Q'_n}} = {\w_2}_{\bound{{\tau_2}}{Q'_n}}$, from
  which the ``if'' part of Property (\ref{it:passoc}) follows. The
  ``only if'' part follows by reasoning in the reverse direction.
  Indeed, if $\w \in A$, $\w \in \pi(I)$, so $\w \in C$ and we
  explained earlier that a run $\tau$ of $A$ on $\w$ gives us an
  accepting run of $A_{T_1}$ on $\w_1$ and an accepting run of $A_2$
  on $\w_2$ with $\w = \w_1 \w_2$ and $\w = \w_1 \cm \w_2$ in $I$.
  Since the states of $A_{T_1}$ and $A_2$ are disjunct, the induction
  hypothesis can be applied, immediately giving the desired result.

  Property (\ref{it:pbound}) immediately follows from the induction
  hypothesis if $n \not = \emptyw$, and from the remarks made about
  the accepting run of $A$ otherwise.
  
  \paragraph{$\pat = (\pat_1 \con \pat_2) \con \pat_3$}
  Let $\pat' = \pat_1 \con (\pat_2 \con \pat_3)$, $(A',f') = H(\pat' ,
  C)$ and $f'(n) = (Q'_n, I'_n, F'_n)$. By construction, $I_1 = (Q'_{1}
  \cup Q'_{21}) \cap I_{A'}$ and $F_1$ contains those nodes of
  $F'_{21}$ from which there is a $q \in F'_{22}$ such that
  $\delta_{A'}(q,\emptyw, q')$ and such that there is a path from $q'$
  to a state in $F_{A'}$. 
  
  We will first prove Property (\ref{it:passoc}). Let $\w \in C$ and
  $\match{\w}{\pat}{V}$. In the proof of Proposition \ref{prop:concon}
  we noted that this holds iff $\match{\w}{\pat'}{V'}$ with
  $V'(\emptyw) = V(\emptyw)$, $V'(1)V'(21) = V(1)$, $V'(1n) = V(11n)$,
  $V'(21n) = V(12n)$ and $V'(22) = V(2)$. Furthermore, for $n \in
  \{\emptyw, 1, 2, 21,22\}$, $V'(n) \not = \ndef$. By the induction
  hypothesis, we have an accepting run of $\tau$ of $A'$ on $\w'$ with
  $V'(n) = \w_{\bound{\tau}{Q'_n}}$. The ``if'' part then follows by
  construction since $ V(1) = \w_{\bound{\tau}{Q_1}} =
  \bound{\tau}{Q'_1} \bound{\tau}{Q'_{21}} = V'(1) V'(21)$ .  The ``only
  if'' part can be proven by similar arguments, reasoning in reverse.
  
  Property (\ref{it:ptype}) follows from the induction hypothesis and
  Proposition \ref{prop:concon} if $n \not = 1$. Let $B = (Q_1, I_1,
  F_1, \delta_{A'})$ and suppose $\w \in \typeof{1}{\pat}{C}$, i.e.
  there exists some $\w' \in C$ with $\match{\w'}{\pat}{V}$ and $V(1)
  = \w$. We know that there exist $\w_1$, $\w_2$ and $\w_3$ such that
  $\w = \w_1 \w_2$, $V(11) = \w_1$, $V(12) = \w_2$ and $V(2) = \w_3$.
  Equally, $\match{\w'}{\pat'}{V'}$ with $V'(1) = \w_1$, $V'(21) =
  \w_2$ and $V'(22) = \w_3$. Then there is a run $\tau$ of $A'$ on
  $\w'$ with $\w'_{\bound{\tau}{Q'_1}} = \w_1$,
  $\w'_{\bound{\tau}{Q'_{21}}} = \w_2$ and
  $\w'_{\bound{\tau}{Q'_{22}}} = \w_3$. Let $\bound{\tau}{Q'_{21}} =
  (i_1, j_1)$ and $\bound{\tau}{Q'_{22}} = (i_2, j_2)$ By the
  induction hypothesis of (\ref{it:pbound}) on $n = 2$, $j_1 = i_2$
  and we can find $q_1 \in F'_{21}$ and $q_2 \in I'_{22}$ such that
  $(q_1, j_1), \dots, (q_2, i_1)$ occurs in $\tau$.  Necessarily, $q_1
  \in F_1$, so $\w \in L(B)$.  On the other hand, suppose $\w \in
  L(B)$. Let $\tau_1 = (q_0, k_0), \dots, (q_l, k_l)$ be an accepting
  run of B on $\w$. Then there is some $q_{l+1} \in I_{2}$ such that
  $\delta_{A'}(q_l, \emptyw, q_{l+1})$ and such that there is a path
  from $q_{l+1}$ to a state in $F_{A'}$. Let $\w_3$ be the word
  spelled on this path. We can then construct $\tau_2 = (q_{l+1},
  k_{l+1}), \dots, (q_m, k_m)$ such that $\tau_1, \tau_2$ forms the
  accepting run of $A'$ on $\w \w_3$. Obviously, $\min(\pos{\tau}
  {Q_{1}}) = 0$, $\max(\pos{\tau}{Q_{2}}) = \width{\w \w_3}$, $k_l \in
  \pos{\tau}{Q_{1}})$ and $k_{l+1} \in \pos{\tau}{Q_{2}}$. Is it
  possible that $\max(\pos{\tau}{Q_{1}}) > \width{\w} = k_l$ or that
  $\min(\pos{\tau}{Q_{2}}) < \width{\w} = k_{l+1}$?  Suppose it is,
  then we would have $\match{\w \w_3}{\pat}{V}$ with $V(1) = \sigma_1
  \dots \sigma_{j_1}$ and $V(2) = \sigma_{j_2} \dots \sigma_n$ with $n
  = \width{\w \w_3}$ and either $j_1 > \width{\w}$ or $j_2 <
  \width{\w}$. Now, necessarily, $\w \w_3 = V(\emptyw)$ which, as we
  noted in the proof of Proposition \ref{prop:concon}, should equal $
  V(1) \con V(2)$.  However, the length of $\sigma_1 \dots
  \sigma_{j_1} \sigma_{j_2} \dots \sigma_n$ is always greater than
  $n$, contradiction. As such $\w \w_3 \in C$ and $\match{\w
    \w_3}{\pat}{V}$ with $V(1) = \w$, hence $\w \in
  \typeof{1}{\pat}{C}$.
  
  Property (\ref{it:pbound}) also follows from these observations
  while Property (\ref{it:punambig}) follows directly from the
  induction hypothesis.

  \paragraph{$\pat = (\pat_1 \cor \pat_2) \con \pat_3$}
  Let $\pat' = \pat_1 \con \pat_3 \cor \pat_2 \con \pat_3$, $(A',f') =
  H(\pat',C)$ and $f'(n) = (Q'_n, I'_n, F'_n)$. Then, $A' = A_1 \cup
  A_2$ where $(A_1,f_1) = H(\pat_1 \con \pat_3, C)$, $(A_2, f_2) =
  H(\pat_2 \con \pat_3, C - L(\pat_1 \con \pat_3))$ and $f'(\emptyw) =
  A$, $f'(1n) = f_1(n)$ and $f'(2n) = f_2(n)$. Property
  (\ref{it:ptype}) follows from the induction hypothesis on $\pat_1
  \con \pat_3$ and $\pat_2 \con \pat_3$, and the combination of
  Propositions \ref{prop:orcon} and \ref{prop:or} if $n \not =
  \emptyw$. For $n = \emptyw$, we showed in the proof of Proposition
  \ref{prop:orcon} that $\match{\w}{\pat}{V}$ iff
  $\match{\w}{\pat}{V'}$. The property then follows from the induction
  hypothesis, since $V(\emptyw) = V'(\emptyw) = \w'$. Property
  (\ref{it:punambig}) follows directly from the induction hypothesis
  and the fact that a word cannot be in $L(\pat_1 \con \pat_3) \cap C$
  and $L(\pat_2 \con \pat_3) \cap (C - L(\pat_1 \con \pat_3)$ at the
  same time.  Property (\ref{it:passoc}) follows from the induction
  hypothesis, and the relation between $V$ and $V'$ when
  $\match{\w}{\pat}{V}$ and $\match{\w}{\pat}{V'}$ we gave in the
  Proof of proposition \ref{prop:orcon}. In the same proof we
  (indirectly) showed that when $\match{\w}{\pat}{V'}$ holds, either
  $V'(11)$ and $V'(12)$ are different from $\ndef$ or $V'(21)$ and
  $V'(22)$ are, but not both.  Property (\ref{it:pbound}) follows from
  this observation and the induction hypothesis.
\end{proof}

\end{document}